\documentclass[reprint,notitlepage,twocolumns,
superscriptaddress,aps,longbibliography]{revtex4-1}
\usepackage[utf8]{inputenc}
\usepackage{braket}
\usepackage{amsmath}
\usepackage{amsthm}
\usepackage{mathrsfs}
\usepackage{mathtools}
\usepackage{amssymb}
\usepackage{dsfont}
\usepackage{braket}
\usepackage{bm}
\DeclareMathOperator{\tr}{tr}
\usepackage{caption}
\usepackage{subcaption}
\usepackage{color}

\begin{document}
\title{Reliable Quantum Certification of Bosonic Code Preparations}
\author{Ya-Dong Wu}
\affiliation{QICI Quantum Information and Computation Initiative, Department of Computer Science,
The University of Hong Kong, Pokfulam Road, Hong Kong}
\date{\today}

\begin{abstract}
    Bosonic fault tolerant quantum computing requires preparations of Bosonic code states like cat states and GKP states with high fidelity and reliable quantum certification of these states. Although many proposals on preparing these states have been developed, few investigation has been done on how to reliably certify these experimental states. In this paper, we develop approaches to certify whether a continuous-variable state falls inside a certain Bosonic qubit code space by detecting a witness using Gaussian measurements.
    Our results can be applied to certification of various cat codes including two-component cat states, four-component cat states, and squeezed two-component cat states as well as Gottesman-Kitaev-Preskill codes. 
    Then we further apply our approach to certify resource states in continuous-variable universal fault tolerant measurement-based quantum computing and quantum outputs of instantaneous quantum polynomial circuits, which can be used to show quantum supremacy. The sample complexity of our approach is efficient in terms of number of modes, so significantly reducing overhead compared to quantum state tomography in certification of many-mode quantum states. 
\end{abstract}

\maketitle

\section{Introduction}
A scalable fault tolerant universal quantum computer demands quantum error correction codes to protect qubit information from decoherence noises~\cite{shor1995,NC10}. One promising type of quantum error correcting codes is Bosonic quantum codes that encode a logical qubit into a quantum harmonic oscillator, such that a two-dimensional code space lies in a corner of an infinite-dimensional Hilbert space. This type of quantum error correcting codes include cat codes~\cite{Cochrane1999,Ralph2003,Leghtas2013,mirrahimi2014,bergmann16}, Gottesman-Kiteav-Preskill (GKP) codes~\cite{GKP2001}, and some other rotational symmetric codes~\cite{michael2016,albert2018,grimsmo2020}. 

Significant experimental progress have been made on encoding qubit information into cat states~\cite{ofek2016} and GKP states~\cite{fluhmann2019,campagne2020} on superconducting cavities as well as trapped ions, along with many theoretical proposals on state preparation and engineering~\cite{terhal2016,puri2017,Weigand2018,shi2019}. Then a natural question is how to experimentally characterize these prepared states in harmonic oscillators.
To implement fault tolerant quantum computing with Bosonic quantum error correcting codes, efficient and reliable quantum certification of scalable Bosonic code state preparations is significant. 

The most common approach to characterize a quantum state in experiments is quantum state tomography~\cite{Lvo09,DAr03}, which, however, requires a huge number of measurements and long classical postprocessing time. 
Quantum state tomography cannot handle multi-mode entangled nonGaussian states.
Quantum tomography with neural networks~\cite{tiunov2020,Ahmed2021PRL,Ahmed2021PRR,zhu2022,wu2022} evidently reduces the required number of measurements and shorten the classical postprocessing time, but still cannot deal with multi-mode quantum states.
In practical applications, physicists have the prior knowledge about the classical description of a target quantum state. Hence, rather than fully characterizing an experimental quantum state, in these scenarios, physicists can apply quantum certification~\cite{Eis20}, to efficiently determine whether a prepared quantum state is close enough to certain target state. 
L.\ Aolita et al. investigated quantum certification of photonic states~\cite{Aol15}, including Gaussian states, and those states prepared in Boson sampling~\cite{aaronson2013} and Knill–Laflamme–Milburn (KLM) schemes~\cite{knill2001}. Later U.\ Chabaud developed another verification protocol of quantum outputs in Boson sampling~\cite{Uly20}.
Up to now, how to certify cat states and GKP states, without performing quantum state tomography or fidelity estimation~\cite{da11,Uly19}, is still open. 
On the other hand, although lots of work have been done on efficient verification of many-qubit states, different nature between CV states and qubit states makes certification approaches of many-qubit states~\cite{pallister2018} cannot be directly applied in CV quantum systems~\cite{liu2021}. 

In this paper, we generalize the concept of fidelity witness~\cite{Aol15,Glu18} to witness of Bosonic code space, and propose protocols to certify both two-component cat code space, four-component cat code space, and a realistic GKP code space with finite truncation in phase space. 
The measurements in all the protocols are experimentally friendly, being either homodyne detections or heterodyne detections. 
We propose certification protocols for both the resource states of CV fault-tolerant measurement-based quantum computing~\cite{men14} and the quantum outputs of CV instantaneous quantum polynomial-time (IQP) circuits~\cite{douce2017} by estimating fidelity witnesses.

\section{Certification of Cat State Code Space}
Let $\mathcal H$ denote the infinite-dimensional Hilbert space of a quantum harmonic oscillator. 
A quantum device outputs $n$ copies of quantum states $\rho^{\otimes n}$ on $\mathcal H^{\otimes n}$. Given a Bosonic code logical subspace $\bar{\mathcal H}\subset \mathcal H$, an experimenter, who has no knowledge of $\rho$, wants to determine whether $\tr_{\bar{\mathcal H}}\rho \ge 1-\epsilon$,  by applying local measurement at each copy of $\rho$. 
To certify whether a single-mode state $\rho$ falls on $\bar{\mathcal H}$, we measure an observable $W$, which satisfies the following conditions: (i) if $\rho$ falls on $\bar{\mathcal H}$, i.e. $\tr_{\bar{\mathcal H}}\rho=1$, then $\braket{W}_{\rho}=1$; (ii) otherwise, $\braket{W}_{\rho}<1$. We call the observable $W$ a witness of the code space $\bar{\mathcal H}$. By estimating $\braket{W}_\rho$, we can determine whether to accept $\rho$ as a reliable Bosonic code state or not. In this subsection, we introduce code witnesses for different cat codes and explain how to estimate their mean values by Gaussian measurements.

A Schrodinger cat state is a superposition of coherent states with opposite amplitudes~\cite{buvzek1995} and can be used to encode qubit information~\cite{Cochrane1999}.
We call this code two-component cat code to differentiate it from the cat code with a superposition of four coherent states.
A two-component cat code space $\bar{\mathcal H}$ is spanned by
\begin{align*}
&\ket{\bar{0}}=   \frac{1}{\sqrt{2(1+e^{-2|\alpha|^2})}}\left(\ket{\alpha}+ \ket{-\alpha}\right)\\
&\ket{\bar{1}}=   \frac{1}{\sqrt{2(1+e^{-2|\alpha|^2})}}\left(\ket{\alpha}- \ket{-\alpha}\right).
\end{align*}
A two-component cat code witness is
\begin{equation}\label{eq:t-catwitness}
    W=\mathds{1}-\frac{\left(\hat{a}^{\dagger 2}-\alpha^{*\,2}\right)\left(\hat{a}^{2}-\alpha^2\right)}{2}.
\end{equation}

The two-component cat code space is the degenerate ground space of the Hamiltonian 
\begin{equation}
    \hat{H}_{tCat}=\left(\hat{a}^{\dagger 2}-\alpha^{*\,2}\right)\left(\hat{a}^{2}-\alpha^2\right)
\end{equation}
with eigenvalue zero.
$\hat{H}$ can be written as
\begin{equation}
\hat{H}_{tCat}=0\left( \ket{\bar{0}}\bra{\bar{0}}_{tCat}+\ket{\bar{1}}\bra{\bar{1}}_{tCat}\right)+\lambda_1 \ket{\psi}\bra{\psi}+\cdots
\end{equation}
where $\ket{\psi}$ is the first excited state of $\hat{H}_{tCat}$ and $\lambda_1$ is the first excited energy.
 Then
 \begin{equation}
 \mathds{1}-\frac{\hat{H}_{tCat}}{2}=\ket{\bar{0}}\bra{\bar{0}}_{tCat}+\ket{\bar{1}}\bra{\bar{1}}_{tCat} +\left(1-\frac{\lambda_1}{2} \right)\ket{\psi}\bra{\psi}-\cdots
 \end{equation}
As $\lambda_1\ge 2$, we get
\begin{equation}
 \mathds{1}-\frac{\hat{H}_{tCat}}{2}\le \ket{\bar{0}}\bra{\bar{0}}_{tCat}+\ket{\bar{1}}\bra{\bar{1}}_{tCat}.
\end{equation}
Furthermore, any state $\rho$ satisfying $\tr\left[\rho \left(\mathds{1}-\frac{\hat{H}_{tCat}}{2}\right)\right]=1$
if and only if $\rho$ falls on the two-component cat code space. 
The witness of squeezed two-component cat code space can be obtained by using the fact that squeezing operation is a Gaussian operation inducing a linear transformation on phase space.

To clarify how to measure this witness, for simplicity, we rewrite it in terms of quadrature operators when $\alpha\in \mathbb R$,
\begin{align}\notag
    W_{tCat}&=\frac{3-2\alpha^4}{4}\mathds{1}-\frac{1}{16}(\hat{x}^4+\hat{p}^4)
     -\frac{1}{12}\Bigg[\left(\frac{\hat{x}+\hat{p}}{\sqrt{2}}\right)^4+ \\ \label{eq:catcodewitnessquadrature}
     & \left(\frac{\hat{x}-\hat{p}}{\sqrt{2}}\right)^4\Bigg]
     +\frac{1}{2}\left( 1+\alpha^2\right)\hat{x}^2
     +\frac{1}{2}\left(1-\alpha^2\right)\hat{p}^2,
\end{align}
where $\hat{x}=\frac{\hat{a}+\hat{a}^\dagger}{\sqrt{2}}$ and $\hat{p}=\frac{\hat{a}-\hat{a}^\dagger}{\sqrt{2}\text{i}}$ are the position and momentum operators.
The expectation value of such a witness can be estimated by homodyne detections on at most four different quadrature bases, i.e. $\hat{x}$, $\hat{p}$, $\frac{\hat{x}+\hat{p}}{\sqrt{2}}$ and $\frac{\hat{x}-\hat{p}}{\sqrt{2}}$. 
The witness (\ref{eq:t-catwitness}) can also be rewritten as
\begin{equation}
    W_{tCat}=-\frac{1}{2}\hat{a}^2\hat{a}^{\dagger\,2}+2\hat{a}\hat{a}^{\dagger}+\frac{1}{2}(\alpha^{* 2}\hat{a}^2+\alpha^2\hat{a}^{\dagger\,2})-\frac{1}{2}|\alpha|^4.
\end{equation}
Each term in the above expression is a product of annihilation operators and creation operators in anti-normal order. Using optical equivalence principle~\cite{scully1997}, i.e.,
\begin{equation}
    \braket{\hat{a}^m\hat{a}^{\dagger\, n}}=\int \text{d}^2 \alpha Q_\rho(\alpha) \alpha^m\alpha^{* n} , \quad m,n\in \mathbb N
\end{equation}
$\braket{W_{tCat}}_\rho$ can be estimated by sampling Humusi-Q function $Q_\rho(\alpha):=\braket{\alpha|\rho|\alpha}$ using heterodyne detections.

Four-component cat code, which is superposition of four coherent states, is introduced to protect qubit information from photon loss errors~\cite{mirrahimi2014}. 
A four-component cat code space $\bar{\mathcal H}$ with even parity is spanned by
\begin{align*}
    &\bar{\ket{0}}=\frac{1}{\sqrt{2(1+e^{-2|\alpha|^2})}}\left(\ket{\alpha}+ \ket{-\alpha}\right),   \\
    &\bar{\ket{1}}=\frac{1}{\sqrt{2(1+e^{-2|\alpha|^2})}}\left(\ket{\text{i}\alpha}+ \ket{-\text{i}\alpha}\right). 
\end{align*}
To certify whether a single-mode state $\rho$ falls on $\bar{\mathcal H}$, we measure the witness
\begin{equation}\label{eq:fourCat}
    W_{fCat}=\frac{\mathds{1}+(-1)^{\hat{n}}}{2}-\frac{\hat{H}_{fCat}}{24},
\end{equation}
where $\hat{n}=\hat{a}^\dagger\hat{a}$ is photon number operator, and  $
\hat{H}_{fCat}:=\left(\hat{a}^{\dagger 2}+\alpha^{* 2}\right)\left(\hat{a}^{\dagger 2}-\alpha^{* 2}\right)\left(\hat{a}^{2}-\alpha^2\right)\left(\hat{a}^{2}+\alpha^2\right)$.
The first term is the parity of photon number, equal to the value of Wigner function at origin, and can be measured by a parity measurement.
The second term can be written as
\begin{align}\notag
    \frac{\hat{H}_{fCat}}{24}=&\hat{a}^4\hat{a}^{\dagger\,4}/24-2/3\hat{a}^3\hat{a}^{\dagger\, 3}+3\hat{a}^2\hat{a}^{\dagger\,2}-4\hat{a}\hat{a}^\dagger-\alpha^4/24\hat{a}^{\dagger\,4}\\ \label{eq:fourcomponentCatwitnessantinormal}
    &-\alpha^{*\, 4}/24\hat{a}^4+(|\alpha|^8/24+1)\mathds{1}.
\end{align}
Again the expectation value of each term can be estimated by heterodyne detections.

A squeezed two-component cat code~\cite{schlegel2022} consists of superpositions of two squeezed coherent states
\begin{align*}
&\ket{\bar{0}}\propto \left(D(-\alpha)+ D(\alpha)\right)S(r)\ket{0}\\
&\ket{\bar{1}}\propto \left(D(-\alpha)- D(\alpha)\right)S(r)\ket{0},
\end{align*}
where $S(r):= \text{e}^{1/2(r^*\hat{a}^2-r\hat{a}^{\dagger\, 2})}$ is a squeezing operation, $r\in \mathbb C$ is a squeezing parameter, and $\text{arg}(r)=\text{arg}(\alpha)/2$ such that the squeezed quadrature is in the same direction as that of the displacement operation. 

For simplicity, we suppose $\alpha\in\mathbb R$. 
From the fact that $D(\alpha)S(r)=S(r)D(\alpha \text{e}^{r})$, we obtain a witness of the squeezed two-component cat code space
\begin{equation}\label{eq:witnessSqueezedCat}
    W_{stCat}=\mathds{1}-\frac{S(r) \left(\hat{a}^{\dagger 2}-\alpha^{2}\text{e}^{2r}\right)\left(\hat{a}^{2}-\alpha^2\text{e}^{2r}\right)S(r)^\dagger}{2}.
\end{equation}
Using the fact that $S(r)\hat{x}S(r)^\dagger=\text{e}^{-r}\hat{x}$ and $S(r)\hat{p}S(r)^\dagger=\text{e}^{r}\hat{p}$, the witness~(\ref{eq:witnessSqueezedCat}) can be rewritten in terms of quadrature operators
\begin{align*}
    W_{stCat}=&\frac{3-2\alpha^4 \text{e}^{4r}}{4}\mathds{1}-\frac{1}{16}\left(\text{e}^{-4r}\hat{x}^4+\text{e}^{4r}\hat{p}^4\right)\\
     &-\frac{1}{48}\left[\left(\text{e}^{-r}\hat{x}+\text{e}^{r}\hat{p}\right)^4 + \left(\text{e}^{-r}\hat{x}-\text{e}^{r}\hat{p}\right)^4\right]\\
     &+\frac{1}{2}\left( \text{e}^{-2r}+\alpha^2 \right)\hat{x}^2+\frac{1}{2}\left( 1-\alpha^2\text{e}^{2r}\right)\text{e}^{2r}\hat{p}^2,
\end{align*}
which can be estimated by applying homodyne detections at four quadratures.

\section{Certification of a realistic GKP state}

GKP states are defined to be simultaneous eigenstates of two commutative displacement operators $S_q=e^{2\text{i}\sqrt{\pi} \hat{x}}$ and $S_p=e^{-2\text{i} \sqrt{\pi} \hat{p}}$, with eigenvalue one~\cite{GKP2001}. 
A qubit is encoded into a GKP state in the following way
 \begin{align}
 \ket{\bar{0}}_{iGKP}&:=\sum_{k=-\infty}^{\infty} \delta(x-2k\sqrt{\pi})\ket{x}, \\
  \ket{\bar{1}}_{iGKP}&:=\sum_{k=-\infty}^{\infty} \delta\left(x-(2k+1)\sqrt{\pi}\right)\ket{x}.
 \end{align}
Ideal GKP states are coherent superpositions of position eigenstates, demanding infinite squeezing, which is not physically realistic. 
 From now on, we consider realistic GKP states with finite squeezing.

 A realistic GKP state replaces position eigenstates by finitely squeezed states, i.e.,
  \begin{align}
 \ket{\bar{0}}_{rGKP}&\propto \sum_{k=-\infty}^{\infty} e^{- 4k^2\sigma^2 \pi } D(2k\sqrt{\pi}) \ket{\psi_x}, \\
  \ket{\bar{1}}_{rGKP}&\propto \sum_{k=-\infty}^{\infty} e^{- (2k+1)^2\sigma^2 \pi} D((2k+1)\sqrt{\pi}) \ket{\psi_x},
 \end{align}
 where $D(\alpha):=\text{e}^{\text{i}\sqrt{2}(-\text{Re}(\alpha)\hat{p}+\text{Im}(\alpha)\hat{x})}$ is a displacement operator, $\ket{\psi_x}=\frac{1}{\sqrt{\sigma}\pi^{1/4}}\int \text{d}x \text{e}^{-\frac{x^2}{2\sigma^2}}\ket{x}$ is a position-squeezed vacuum state, and $0<\sigma<1$ is the variance of a Gaussian distribution.
When $\sigma$ is small, a GKP state that is a superposition of Gaussian peaks with width~$\sigma$, separated by~$2\sqrt{\pi}$, with Gaussian envelop of width~$1/\sigma$ in position basis is
 a superposition of Gaussian peaks with width~$\sigma$, separated by~$\sqrt{\pi}$, with Gaussian envelop of width~$\frac{1}{\sigma }$ in momentum basis.

 A straightforward way to certify a GKP state is to detect the eigenvalue of two stabilizer operators $S_q$ and $S_p$ to check whether both eigenvalues are one. 
 However, a realistic GKP state with finite squeezing can never pass this certification test. 
 To circumvent this obstacle, one solution is to set a threshold on the eigenvalue in the certification test to pass those realistic GKP states which are close enough to the ideal GKP states.
 Nevertheless, this approach does not exclusively certify a realistic GKP state or a GKP code space with a certain target degree of squeezing.
 In this paper, we aim to certify a realistic GKP state with a certain target degree of squeezing.

To certify a realistic GKP state, we must truncate the phase space to consider only finite number of superpositions of squeezed coherent states. Suppose we truncate at $x=\pm\sqrt{\pi}m$ in position quadrature and $p=\pm\sqrt{\pi}m$ in momentum quadrature, respectively, where $m\in\mathbb N^+$. 
Then we obtain a GKP code space witness 
 \begin{align}\notag
&W_{rGKP}=\\ \notag
&\mathds{1}-\frac{1}{(2m+1)!} \prod_{-m\le k\le m}\left(\cosh r\hat{a}^\dagger-\sinh r\hat{a} -\frac{\sqrt{\pi }k}{\sigma}\right)\cdot\text{h.c.}\\ \label{eq:GKPcode}
&-\frac{1}{(2m+1)!} \prod_{-m\le k\le m}\left(\cosh r\hat{a}^\dagger+\sinh r\hat{a}+\text{i}\frac{\sqrt{\pi }k}{\sigma}\right)\cdot \text{h.c.},
 \end{align}
 where $r=\frac{1}{2}\ln\frac{1}{\sigma}$ and $\text{h.c.}$ denotes Hermitian conjugate. 
 As GKP states are applied in fault-tolerant CV quantum computing, we use this witness to certify output state in fault-tolerant CV quantum computing.

\section{Application in verification of fault-tolerant quantum computing}
\begin{figure}
    \centering
    \includegraphics[width=0.45\textwidth]{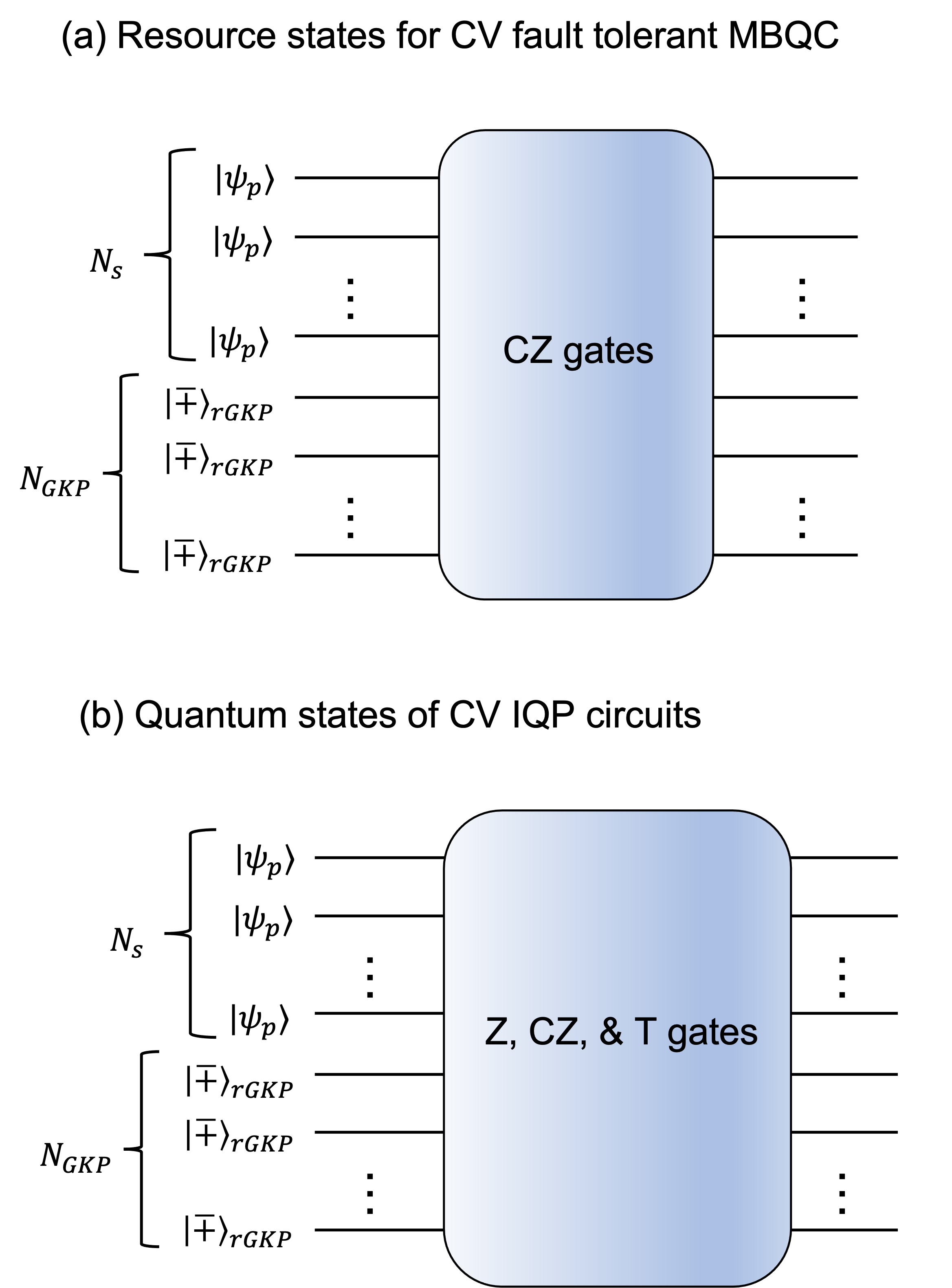}
    \caption{Diagrams of quantum circuits to generate the two types of target states. }
    \label{fig:circuits}
\end{figure}

In this subsection, we apply the results in the last subsection to certify two important types of many-mode quantum states. 
The first are the resource states in universal fault-tolerant CV measurement-based quantum computing~\cite{men14}. These states are CV cluster states~\cite{men06,Mil09,larsen2019,hastrup2021}, attached with GKP states. The second are the output states of CV IQP circuits~\cite{douce2017}. These states are prepared by applying unitary gates diagonal in position quadrature on the combination of momentum-squeezed states and GKP states. The preparations of both two types of target states are plotted in Fig.~\ref{fig:circuits}.

 We first slightly revise the code witness in Eq.~(\ref{eq:GKPcode}) to obtain a fidelity witness of the GKP state 
 \begin{equation}
 \ket{\bar{+}}_{rGKP}\propto \mathcal N_0 \sum_{k=-\infty}^{\infty} e^{- \sigma^2 \pi k^2} D(\sqrt{\pi}k) \ket{\psi_x}.
 \end{equation}
 As $\ket{\bar{+}}_{rGKP}$ is a grid of Gaussian peaks separated by $\sqrt{\pi}$ in position quadrature and separated by $2\sqrt{\pi}$ in momentum quadrature. 
 We obtain a fidelity witness 
 \begin{align}\notag
&W_{\ket{\bar{+}}_{rGKP}}=\mathds{1}-\\ \notag
&\frac{1}{(2m+1)!} \prod_{-m\le k\le m}\left(\cosh r\hat{a}^\dagger-\sinh r\hat{a} -\frac{\sqrt{\pi }k}{\sigma}\right)\cdot\text{h.c.}- \\ \label{eq:GKPstate}
&\frac{1}{(m+1)!} \prod_{-m/2\le k\le m/2}\left(\cosh r\hat{a}^\dagger+\sinh r\hat{a}+2\text{i}\frac{\sqrt{\pi }k}{\sigma}\right)\cdot \text{h.c.}
 \end{align}

 
A CV cluster state can be considered as more generally a CV graph state, where each vertex represents a quantum mode and each edge connecting vertices $i$ and $j$ represents a quantum gate $CZ:=\text{e}^{\text{i}\hat{x}_i\otimes \hat{x}_j}$.
Suppose a target cluster state has $N_s+N_{GKP}$ modes in total, where $N_s$ modes are initially momentum-squeezed vacuum states $\ket{\psi_p}:=\frac{1}{\sqrt{\sigma}\pi^{1/4}}\int \text{d}p \text{e}^{-\frac{p^2}{2\sigma^2}}\ket{p}$ and the other $N_{GKP}$ modes are initially $\ket{\bar{+}}_{rGKP}$. The target CV cluster state is obtained by applying a CZ gate at each pair of adjacent modes $i$ and $j$ in the graph.

A fidelity witness of this resource state is
\begin{widetext}
\begin{align}\notag
W=&\left(1+\frac{N_s}{2}\right)\mathds{1}-\sum_{i=1}^{N_{GKP}}\Bigg[\frac{1}{(2m+1)!} \prod_{-m\le k\le m}\left(\frac{\text{e}^{-r}}{\sqrt{2}}\hat{x}_i-\frac{\text{e}^r}{\sqrt{2}}\text{i}\hat{\tilde{p}}_i-\frac{\sqrt{\pi }k}{\sigma}\right)\cdot h.c.\\ \label{eq:resourceState}
&+\frac{1}{(m+1)!} \prod_{-m/2\le k\le m/2} \left(\frac{\text{e}^{r}}{\sqrt{2}}\hat{x}_i-\frac{\text{e}^{-r}}{\sqrt{2}}\text{i}\hat{\tilde{p}}_i+2\text{i}\frac{\sqrt{\pi }k}{\sigma}\right)\cdot h.c.\Bigg]-\frac{1}{2}\sum_{i=1}^{N_s} \left(e^{2r}\hat{x}_{i}^2 +e^{-2r}\hat{\tilde{p}}_{i}^2\right),
\end{align}
where $\hat{\tilde{p}}_i=\hat{p}_i-\sum_{j\in\mathcal N(i)}\hat{x}_j$ and $\mathcal N(i)$ denotes the set of modes adjacent to mode $i$ in the graph.
\end{widetext}
The fidelity witness in Eq.~(\ref{eq:resourceState}) contains at most $(n+2)^{4m+2}$ products of quadrature operators with maximal order $4m+2$, where $n$ is the maximal number of neighborhood modes and is as most four in a square-lattice cluster state.
To calculate the required sample complexity to estimate the mean value of fidelity witness, we
denote $\sigma_k$ to be the uniform upper bound of the mean square of all products of $k$ quadrature operators and $\sigma_{\le k}$ to be the maximum value of all $\sigma_j$ with $1\le j\le k$.
Suppose $m$ is a constant, that is we always truncate the phase space of each mode at a certain fixed value in both position and momentum quadratures.
If we use the approach of importance sampling~\cite{flammia2011} to estimate $\braket{W}$, 
by Hoeffding's inequality, we find that the minimum number of required copies of states to obtain an estimate $\omega$ such that
$\text{Pr}\left(|\omega-\braket{W}_\rho|\ge\epsilon\right)\le \delta$,
is upper bounded by 
 \begin{equation}
          O\left[\frac{\ln1/\delta}{\epsilon^2}\left(N_{GKP}^2\text{e}^{r(4m+2)}\sigma_{\le 4m+2}+N_s^2\text{e}^{2r}\sigma_{2}\right) \right].
 \end{equation}
This sample complexity scales polynomially in both $N_s$ and $N_{GKP}$.

A CV IQP circuit~\cite{douce2017} is a uniformly random combination of three quantum gates in the set
$$\left\{Z:=e^{\text{i}\hat{x}\sqrt{\pi}}, CZ, 
T:=e^{\text{i}\frac{\pi}{4}[2(\hat{x}/\sqrt{\pi})^3+(\hat{x}/\sqrt{\pi})^2-2\hat{x}/\sqrt{\pi}]}\right\},$$
with the input of combination of $N_s$ copies of $\ket{\psi_p}$ and $N_{GKP}$ copies of $\ket{\bar{+}}_{rGKP}$.
 Denote $n_Z^i$, and $n_{T}^i$ as the number of Z gates and T gates applied at $i$th mode, respectively. 
Then the fidelity witness of the output state of a CV IQP circuit is given in Eq.~(\ref{eq:resourceState}) with
 \begin{equation*}
     \hat{\tilde{p}}_i=\hat{p}_i-n_T^i(3\hat{x}_i^2/(2\sqrt{\pi})+\hat{x}_i/2-\sqrt{\pi}/2)
     -n_Z^i\sqrt{\pi}-\sum_{j\in\mathcal N(i)}\hat{x}_{j}
\end{equation*}

There are at most $N_{GKP}(3+n_{CZ})^{4m+2}$ terms of products of quadrature operators with order at most $4m+2$, and the absolute value of coefficient is no larger than $(n_T\text{e}^r)^{4m+2}$. Thus the sample complexity is  bounded by
\begin{align*}
    O&\Bigg[\frac{\ln1/\delta}{\epsilon^2} \Big(N_{GKP}^2(n_T\text{e}^r)^{4m+2}(n_{CZ}+3)^{8m+4}\sigma_{\le 4m+2}+\\
    &N_s^2 n_T^2 (n_{CZ}+3)^4 \sigma_{\le 4}\Big)\Bigg].
\end{align*}
The sample complexity scales polynomially with respect to both the number of modes and the number of gates.

\section{Methods}
\label{sec:method}

The four-component cat code space is the degenerate ground space of Hamiltonian
\begin{align}\notag
\hat{H}=&\left(\hat{a}^{\dagger 2}+\alpha^{* 2}\right)\left(\hat{a}^{\dagger 2}-\alpha^{* 2}\right)\left(\hat{a}^{2}-\alpha^2\right)\left(\hat{a}^{2}+\alpha^2\right)\\
=&\hat{a}^{\dagger 4}\hat{a}^4-(\alpha^4\hat{a}^{\dagger 4}+\alpha^{* 4}\hat{a}^4)
+|\alpha^2|^4
\end{align}
with even photon parity. 
Following idea similar to above two-component cat code, and using the fact that $\frac{\mathds{1}+(-1)^{\hat{n}}}{2}$ is the projection onto the even parity subspace, 
we obtain the witness of four-component cat code space as shown in Eq.~(\ref{eq:fourCat}). 

A realistic GKP state is a superposition of squeezed coherent states.
 Each component $D(\sqrt{\pi}k)\ket{\psi_x}$ has the nullifier $S(r)\left(\hat{a}^\dagger -\frac{\sqrt{\pi}k }{\sigma}\right)\left( \hat{a} -\frac{\sqrt{\pi}k}{\sigma}\right) S(r)^\dagger$.
Then the superposition of $D(\sqrt{\pi}k) \ket{\psi_x}$ for $k\in \mathbb{Z}$ has the nullifier
 \begin{equation}
S(r) \prod_{k\in \mathbb{Z}}\left(\hat{a}^\dagger -\frac{\sqrt{\pi }k}{\sigma}\right) \prod_{l\in \mathbb{Z}} \left(\hat{a}-\frac{\sqrt{\pi }l}{\sigma}\right) S(r)^\dagger.
 \end{equation}
Similarly, any state $D( 2\sqrt{\pi} k\text{i} ) \ket{\psi_p}$ has the nullifier 
 \begin{equation}
 S(\text{i} r) \left(\hat{a}^\dagger+\text{i}\frac{2\sqrt{\pi} k}{\sigma}\right)  \left(\hat{a}-\text{i}\frac{2\sqrt{\pi} k}{\sigma }\right)  S(\text{i}r)^\dagger.
  \end{equation}
 Thus, a realistic GKP state has another nullifier
  \begin{equation}
  S ( \text{i}r) \prod_{k\in \mathbb{Z}}\left(\hat{a}^\dagger +\text{i}\frac{2\sqrt{\pi} k}{\sigma }\right) \prod_{l\in \mathbb{Z}} \left(\hat{a}-\text{i}\frac{2\sqrt{\pi} l}{\sigma }\right) S(\text{i}r)^\dagger.
  \end{equation}
 Combining these two nullifiers together, we obtain the fidelity witness of a realistic GKP state as shown in Eq.~(\ref{eq:GKPstate}).

To obtain the fidelity witness of a cluster state attached with GKP states, we also need a witness of CV cluster state.
A cluster state is
 constructed from tensor product of $n$ $\hat{p}$-squeezed vacuum states by applying CV CZ gates.
A nullifier of each mode $1\le i \le n$ in a $n$-mode CV cluster state is ,
\begin{equation}
\frac{e^{2r} \left(\hat{x}_{i}^2 - \sum_{j\in \mathcal N(i)} \hat{x}_j^2\right)+e^{-2r}\hat{p}_{i}^2-1}{2}.
\end{equation}
Combining all the nullifiers, we obtain a fidelity witness in Eq.~(\ref{eq:resourceState}).
The fidelity witness of output states of IQP circuits can be obtained  by noticing that 
$$
        T\hat{p}T^\dagger=\hat{p}-3\hat{x}^2/(2\sqrt{\pi})-\hat{x}/2+\sqrt{\pi}/2,
$$ and following a similar strategy.

The sample complexities for both the certification protocols of CV cluster states attached with GKP states and output states of IQP circuits can be calculated using importance sampling and Hoeffding's inequality. Suppose a random variable $\bm{F} =\sum_{i=1}^m \lambda_i f_i$, where each $f_i$ is a polynomial function of quadrature operators. Then by importance sampling approach,
Hoeffindg's inequality implies
$\text{Pr}(|F^*-\bm{F}|>\epsilon)\le 8\text{e}^{-\frac{N\epsilon^2}{33\braket{\bm{F}^2}}}$, where $\braket{\bm{F}^2}\le m^2\max_i|\lambda_i| \max_i \braket{f_i^2}$. As calculated in Ref.~\cite{Liu19,farias2021}, to make the failure probability of estimation less than $\delta$, the required sample complexity is upper bounded by $O\left(\frac{\ln(1/\delta)}{\epsilon^2} m^2\max_i|\lambda_i| \max_i\braket{f_i^2}\right)$.

\section{Discussion}

Bosonic quantum error correcting codes is a promising way to realize universal fault tolerant quantum computing is a qubit-into-qumode manner alternative to the KLM scheme. Preparation of bosonic code quantum states with high fidelity is an key issue for the implementation of Bosonic quantum error correction. Here we propose realistic protocols to certify the preparations of experimental Bosonic code sates using Gaussian measurements. Different from state tomography, this protocol is extended to certification of many-mode quantum states efficiently with respect to the number of modes. Most previous work on certification or verification of nonGaussian states are about nonGaussian states with finite stellar rank~\cite{Chabaud2020prl}. In contrast to quantum states in Boson sampling and KLM schemes, cat states and GKP states have infinite stellar rank, indicating higher nonGaussianity.

In this paper, we first develop an approach to certify whether a CV state falls inside a certain two-dimensional Bosonic code space. 
If a CV state passes this certification protocol, then we can ignore the CV nature of this state and consider only the logical code space.
By combining our certification approach, together with quantum characterization approaches on many-qubit systems, including direct fidelity estimation~\cite{flammia2011}, qubit-state verification~\cite{pallister2018} and classical shadow estimation~\cite{huang2020}, one can certify preparations of many-qubit Bsonic code states.

Although here we assume independent and identical copies of prepared quantum states, we can extend this work into non-i.i.d scenario by using the technique developed in Ref.~\cite{wu2021}. Hence, the protocol can be applied to verifiable blind fault-tolerant quantum computing~\cite{hayashi2015,gheorghiu2019}. A server prepares quantum states, which are claimed to be resource states for CV fault-tolerant quantum computing, and sends them to a client. After receiving these states, the client randomly chooses some of them to perform dimension test and fidelity test. If both tests are passed, then the remaining states can be used for measurement-based fault tolerant quantum computing with homodyne detections. 

Besides verification of resource states for universal quantum computing, we also propose protocols to certify output states of IQP circuits. It has been shown that the statstics of practical homodyne measurements on position quadratures of the quantum output of IQP circuits cannot be efficiently simulated by a classical computer~\cite{douce2017}. Hence, a certified quantum output state of IQP circuits can be used to show quantum supremacy. 

\section{Acknowleadgement}
This work was supported by funding from the Hong Kong Research Grant Council through grants no.\ 17300918 and no.\ 17307520.

\bibliography{refs}

\begin{widetext}
\section{Appendix}

\subsection{Calculation of two-component cat code witness in Eq.~(\ref{eq:catcodewitnessquadrature})}
By using 
\begin{align}
\hat{a}^{\dagger\, 2}\hat{a}^2=& \hat{n}^2-\hat{n}\\
=&\frac{1}{4}\left (\hat{x}^4+\hat{p}^4+\hat{x}^2\hat{p}^2+\hat{p}^2\hat{x}^2\right)-\hat{x}^2-\hat{p}^2+\frac{3}{4} \\
\alpha^{* 2}\hat{a}^2+\alpha^2\hat{a}^{\dagger \, 2}=&\operatorname{Re}(\alpha^2)(\hat{x}^2-\hat{p}^2)+\operatorname{Im}(\alpha^2)(\hat{x}\hat{p}+\hat{p}\hat{x}),
\end{align}
we obtain
\begin{align}
 W=&\mathds{1}-\frac{\hat{H}_{tCat}}{2}\\
 =&\mathds{1}-\frac{1}{8}\left (\hat{x}^4+\hat{p}^4+\hat{x}^2\hat{p}^2+\hat{p}^2\hat{x}^2\right) +\frac{1}{2}(\hat{x}^2+\hat{p}^2)-\frac{3}{8}+\frac{1}{2}\operatorname{Re}(\alpha^2)(\hat{x}^2-\hat{p}^2)\\
 &+\frac{1}{2}\operatorname{Im}(\alpha^2)(\hat{x}\hat{p}+\hat{p}\hat{x})-\frac{|\alpha|^4}{2}
\end{align}

Note that
\begin{equation}
\hat{x}\hat{p}+\hat{p}\hat{x}=(\hat{x}+\hat{p})^2-\hat{x}^2-\hat{p}^2
\end{equation}
and
\begin{equation}
\hat{x}^2\hat{p}^2+\hat{p}^2\hat{x}^2=\frac{1}{6}\left[\left(\hat{x}+\hat{p}\right)^4+\left(\hat{x}-\hat{p}\right)^4\right]-\frac{1}{3}(\hat{x}^4+\hat{p}^4)-1.
\end{equation}
The second equality is because 
\begin{align*}
    (\hat{x}+\hat{p})^4+(\hat{x}-\hat{p})^4=&2(\hat{x}^4+\hat{p}^4+\hat{x}^2\hat{p}^2+\hat{p}^2\hat{x}^2)+2(\hat{x}\hat{p}\hat{x}\hat{p}+\hat{x}\hat{p}^2\hat{x}+\hat{p}\hat{x}\hat{p}\hat{x}+\hat{p}\hat{x}^2\hat{p})\\
    =&2(\hat{x}^4+\hat{p}^4+\hat{x}^2\hat{p}^2+\hat{p}^2\hat{x}^2)+4(\hat{x}^2\hat{p}^2+\hat{p}^2\hat{x}^2)+6.
\end{align*}
So the fidelity witness can be rewritten as
\begin{align}
W=&\mathds{1}-\frac{1}{8}(\hat{x}^4+\hat{p}^4)-\frac{1}{48}\left[ \left(\hat{x}+\hat{p}\right)^4 + \left(\hat{x}-\hat{p}\right)^4\right]+\frac{1}{24}(\hat{x}^4+\hat{p}^4)\\
&+\frac{1}{8}+\frac{1}{2}(\hat{x}^2+\hat{p}^2)-\frac{3}{8}+\frac{1}{2}\operatorname{Re}(\alpha^2)(\hat{x}^2-\hat{p}^2)+\frac{1}{2}\operatorname{Im}(\alpha^2) \left(\hat{x}+\hat{p}\right)^2\\
&-\frac{1}{2}\operatorname{Im}(\alpha^2)\hat{x}^2 -\frac{1}{2}\operatorname{Im}(\alpha^2)\hat{p}^2 -\frac{|\alpha|^4}{2} \\
=&\mathds{1}-\frac{1}{16}(\hat{x}^4+\hat{p}^4)-\frac{1}{48}\left[ \left(\hat{x}+\hat{p}\right)^4 + \left(\hat{x}-\hat{p}\right)^4\right]\\
&+\frac{1}{2}\operatorname{Im}(\alpha^2) \left(\hat{x}+\hat{p}\right)^2+\frac{1}{2}\left[ 1+\operatorname{Re}(\alpha^2)- \operatorname{Im}(\alpha^2)\right]\hat{x}^2\\
&+\frac{1}{2}\left[ 1-\operatorname{Re}(\alpha^2)- \operatorname{Im}(\alpha^2)\right]\hat{p}^2-\frac{1+2|\alpha|^4}{4}.
\end{align}

In the case that $\alpha^2\in \mathbb{R}$, we have
\begin{equation}
W=\mathds{1}-\frac{1}{16}(\hat{x}^4+\hat{p}^4)-\frac{1}{48}\left[ \left(\hat{x}+\hat{p}\right)^4 + \left(\hat{x}-\hat{p}\right)^4\right]+\frac{1}{2}( 1+\alpha^2)\hat{x}^2+\frac{1}{2} (1-\alpha^2)\hat{p}^2-\frac{1+2 \alpha^4}{4}.
\end{equation}

\subsection{Calculation of four-component cat code witness in Eq.~(\ref{eq:fourcomponentCatwitnessantinormal})}

Using the fact that for any $k\in\mathbb N^+$,
\begin{equation}
    \hat{a}^{\dagger k}\hat{a}^k=\hat{n}(\hat{n}-1)\cdots(\hat{n}-k+1)
\end{equation}
and
\begin{align*}
    &\hat{n}=\hat{a}\hat{a}^\dagger-1\\
    &\hat{n}^2=\hat{a}^2\hat{a}^{\dagger\,2}-3\hat{a}\hat{a}^\dagger+1\\
    &\hat{n}^3=\hat{a}^3\hat{a}^{\dagger\,3}-6\hat{a}^2\hat{a}^{\dagger\,2}+7\hat{a}\hat{a}^\dagger-1\\
    &\hat{n}^4=\hat{a}^4\hat{a}^{\dagger\,4}-10\hat{a}^3\hat{a}^{\dagger\,3}+25\hat{a}^2\hat{a}^{\dagger\,2}-15\hat{a}\hat{a}^\dagger+1
\end{align*}
we get
\begin{align*}
    &\hat{a}^{\dagger 4}\hat{a}^4=\hat{a}^4\hat{a}^{\dagger\,4}-16\hat{a}^3\hat{a}^{\dagger\,3}+72\hat{a}^2\hat{a}^{\dagger\,2}-96\hat{a}\hat{a}^\dagger+24,\\
    &\hat{a}^{\dagger 3}\hat{a}^3=\hat{a}^3\hat{a}^{\dagger\,3}-9\hat{a}^2\hat{a}^{\dagger\,2}+18\hat{a}\hat{a}^\dagger-6,\\
    &\hat{a}^{\dagger 2}\hat{a}^2=\hat{a}^2\hat{a}^{\dagger\,2}-4\hat{a}\hat{a}^\dagger+2,\\
    &\hat{a}^\dagger \hat{a}=\hat{a}\hat{a}^\dagger -1.
\end{align*}
Combing all the above equations, we obtain Eq.~(\ref{eq:fourcomponentCatwitnessantinormal}).

\subsection{Calculation of state witness in fault-tolerant quantum computing}
Using the fact that $CZ\hat{a}_i CZ^\dagger= \hat{a}_i-\frac{\text{i}}{\sqrt{2}}\hat{x}_j$, we have
\begin{align*}
    &\prod_{j\in\mathcal N(i)} CZ_{ij}\left(\cosh r\hat{a}_i^\dagger -\sinh r \hat{a}_i-\frac{\sqrt{\pi }k}{\sigma}\right)\prod_{j\in\mathcal N(i)}CZ_{ij}^\dagger\\
    = & \cosh r \hat{a}_i^\dagger +\frac{\text{i}}{\sqrt{2}}\cosh r\sum_{j\in \mathcal N(i)}\hat{x}_j -\sinh r\hat{a}_i+\frac{\text{i}}{\sqrt{2}}\sinh r\sum_{j\in\mathcal N(i)}\hat{x}_j-\frac{\sqrt{\pi }k}{\sigma}\\
    =&\frac{\text{e}^{-r}}{\sqrt{2}}\hat{x}_i-\frac{\text{e}^r}{\sqrt{2}}\text{i}\hat{p}_i+\frac{\text{i}}{\sqrt{2}}\text{e}^r\sum_{j\in\mathcal N(i)}\hat{x}_j-\frac{\sqrt{\pi }k}{\sigma}
\end{align*}
and
\begin{align*}
    &\prod_{j\in\mathcal N(i)}CZ_{ij}\left(\cosh r\hat{a}_i^\dagger  +\sinh r \hat{a}_i+2\text{i}\frac{\sqrt{\pi }k}{\sigma} \right)\prod_{j\in\mathcal N(i)}CZ_{ij}^\dagger\\
    \rightarrow & \cosh r \hat{a}_i^\dagger +\frac{\text{i}}{\sqrt{2}}\cosh r\sum_{j\in \mathcal N(i)}\hat{x}_j +\sinh r\hat{a}_i-\frac{\text{i}}{\sqrt{2}}\sinh r\sum_{j\in\mathcal N(i)}\hat{x}_j+2\text{i}\frac{\sqrt{\pi }k}{\sigma}\\
    =&\frac{\text{e}^{r}}{\sqrt{2}}\hat{x}_i-\frac{\text{e}^{-r}}{\sqrt{2}}\text{i}\hat{p}_i+\frac{\text{i}}{\sqrt{2}}\text{e}^{-r}\sum_{j\in\mathcal N(i)}\hat{x}_j+2\text{i}\frac{\sqrt{\pi }k}{\sigma}.
\end{align*}
Plugging them into Eq.~(\ref{eq:GKPstate}), we obtain the witness in Eq.~(\ref{eq:resourceState}).

Since
\begin{align*}
    &Z\hat{x}Z^\dagger=CZ\hat{x}CZ^\dagger=T\hat{x}T^\dagger=\hat{x}\\
    &Z\hat{p}Z^\dagger=\hat{p}-\sqrt{\pi}\\
    &CZ\hat{p}_1 CZ^\dagger=\hat{p}_1-\hat{x}_2\\
    &T\hat{p}T^\dagger=\hat{p}+\text{i}\pi/4(2/\sqrt{\pi^3}[\hat{x}^3,\hat{p}]+1/\pi [\hat{x}^2,\hat{p}]-2\text{i}/\sqrt{\pi})
    =\hat{p}-3\hat{x}^2/(2\sqrt{\pi})-\hat{x}/2+\sqrt{\pi}/2,
\end{align*}
for each mode in the QIP circuit, the annihilation operator is transformed to
\begin{equation}
    \hat{a}\rightarrow \hat{a}-\text{i}\sqrt{\frac{\pi}{2}}n_Z-\frac{\text{i}}{\sqrt{2}}\sum_{j=1}^{n_{CZ}}\hat{x}_j-3\text{i}n_T \hat{x}^2/(2\sqrt{2}\pi)-\text{i}n_T\hat{x}/2\sqrt{2}+\sqrt{\pi}n_T\text{i}/2\sqrt{2}.
\end{equation}
Hence,
\begin{align*}
    &\cosh r\hat{a}^\dagger-\sinh r\hat{a}\\
    \rightarrow & \cosh r( \hat{a}^\dagger+\text{i}\sqrt{\frac{\pi}{2}}n_Z+\frac{\text{i}}{\sqrt{2}}\sum_{j=1}^{n_{CZ}}\hat{x}_j+3\text{i}n_T \hat{x}^2/(2\sqrt{2}\pi)+\text{i}n_T\hat{x}/2\sqrt{2}-\sqrt{\pi}n_T\text{i}/2\sqrt{2})\\
    &-\sinh r (\hat{a}-\text{i}\sqrt{\frac{\pi}{2}}n_Z-\frac{\text{i}}{\sqrt{2}}\sum_{j=1}^{n_{CZ}}\hat{x}_j-3\text{i}n_T \hat{x}^2/(2\sqrt{2}\pi)-\text{i}n_T\hat{x}/2\sqrt{2}+\sqrt{\pi}n_T\text{i}/2\sqrt{2})\\
    =&(\frac{\text{e}^{-r}}{\sqrt{2}}+\frac{n_T\text{i}\text{e}^r}{2\sqrt{2}})\hat{x}_i-\frac{\text{e}^r}{\sqrt{2}}\text{i}\hat{p}_i+\frac{\text{i}}{\sqrt{2}}\text{e}^r\sum_{j=1}^{n_{CZ}}\hat{x}_j+\frac{3n_T\text{i}\text{e}^r}{2\sqrt{2}\pi}\hat{x}_i^2+\text{i}\sqrt{\frac{\pi}{2}}(n_Z-n_T/2)\text{e}^r.
\end{align*}
Plugging it into Eq.~(\ref{eq:GKPstate}), we obtain the witness of quantum output of IQP circuit.

\end{widetext}
\end{document}